\documentclass[twocolumn]{aa} 

\usepackage{graphicx}
\usepackage{slashbox}
\usepackage{amsmath}
\usepackage{color}
\usepackage{mathptmx} 
\usepackage{multirow}
\usepackage{wasysym}
\usepackage{url}
\usepackage{txfonts}
\usepackage[authoryear]{natbib}

   \title{Asteroid flux towards circumprimary habitable zones in binary star systems: I. Statistical overview}
   \author{D. Bancelin
          \inst{1,2}
          \and E. Pilat-Lohinger\inst{1}
          \and S. Eggl\inst{2}
          \and T.I. Maindl\inst{1}
          \and C. Sch\"afer\inst{3}
          \and R. Speith\inst{4}
          \and R. Dvorak\inst{1}
          }

   \institute{Institute of Astrophysics (ifA), University of Vienna, T\"urkenschanzstr. 17, 1180 Vienna,
Austria (\email{david.bancelin@univie.ac.at})
 \and
 IMCCE, Paris Observatory, UPMC, CNRS, UMR8028, 77, Av. Denfert-Rochereau 75014 Paris,
France
\and 
Institut f\"ur Astronomie und Astrophysik, Eberhard Karls Universit\"at T\"ubingen, Auf der
Morgenstelle 10, 72076 T\"ubingen, Germany 
\and Physikalisches Institut, Eberhard Karls Universit\"at T\"ubingen, Auf der
Morgenstelle 14, 72076 T\"ubingen, Germany 
\\
             }

   \date{Received ---, 2015; accepted ---}

\definecolor{gray}{gray}{0.62}

\begin{document}
%


  \abstract
   {So far, multiple stellar systems harbor more than 130 extra solar planets. Dynamical simulations show that 
the outcome of
planetary formation process can lead to various planetary architecture (i.e. location, size, mass and water content) 
when the star system is single or double.}
   {In the late phase of planetary formation, when embryo-sized objects dominate the inner region of the 
system, asteroids are also present and can provide additional material for objects inside the habitable zone (hereafter 
HZ). In this study, we make a comparison of several binary star systems and their efficiency to move icy asteroids 
from beyond the snow-line into orbits crossing the HZ.}
   {We modeled a belt of 10000 asteroids (remnants from the late phase of planetary formation process) beyond the 
snow-line. The planetesimals are placed randomly around the primary star and move under the gravitational influence of 
the two stars and a gas giant. As the planetesimals do not interact with each other, we divided the belt into 100 
subrings which were separately integrated. In this statistical study, several double star configurations with a 
G-type star as primary are investigated.}
   {Our results show that small bodies also participate in bearing a non-negligible amount of water to the HZ. The 
proximity of a companion moving on an eccentric orbit increases the flux of
asteroids to the HZ, which could result into a more efficient water transport on a short timescale, causing a heavy
bombardment. In contrast to asteroids moving under the gravitational perturbations of one G-type star and 
a gas giant, we show that the presence of a companion star can not only favor a faster depletion of our disk of 
planetesimals but can also bring 4 -- 5 times more water into the whole HZ.}
   {}
   \titlerunning{Asteroids flux to HZ in binary star systems} 
   \maketitle

%

\section{Introduction}

Nearly 130 extra solar planets in double and multiple star systems have been discovered to date.
Roughly one quarter of these planets is orbiting close to or even crossing their system's habitable zone (HZ), i.e. the
region where an Earth-analog
could retain liquid water on its surface \citep{rein14}.
While most of these planets are gas giants, the incredible ratio of one in four planets to be at least partly in the HZ
seems to make binary star systems promising targets in the search of a second Earth,
especially for the next generation of photometry missions CHEOPS, TESS and PLATO-2.0.
About 80 percent \citep{rein14} of the currently known planets in double star systems are in so-called S-type
configurations \citep{dvorak84}, 
i.e. the two stars are so far apart so that the planet orbits only one stellar component without being
destabilized. 
As most of the wide binary systems host not only one gas giant, their dynamical evolution is quite complex. 
 The question whether habitable worlds 
can actually exist in such environments is, therefore, not a trivial one. Previous works on early stages of 
planetary formation show that planetesimals accretion can be more difficult than in single star systems \citep[and 
references therein]{thebault14}. This in turn can question the possibility of forming embryos in such systems. However, 
studies of late stages of planetary formation show that, should embryos manage to form despite these 
adverse conditions, then the dynamical influence of companion stars is 
not 
prohibitive to form  
Earth-like planets \citep{raymond04,haghighipour07}. Furthermore, it was shown that binary star
systems in the
vicinity of the Solar System are capable of sustaining habitable worlds, once they are formed \citep{eggl13,jaime14}.
As the amount of water on a planet's surface seems to be crucial to sustain a temperate environment
\citep{kasting93}, it is important to identify possible sources.
For Earth, two mechanisms seem to be important: i) endogenous outgassing of primitive material and ii) exogenous impact
by asteroids and comets sources have
been established.
Since neither can explain the amount and isotope composition of Earth's oceans in itself, however, models that favor a
combination 
of both sources seem to be more successful \citep{izidoro13}. 
The amount of primordial water that is collected during formation phases of planets in S-type orbits in binary 
star systems containing
additional gas giants has been studied by \cite{haghighipour07}.
They have shown that the planets formed in a circumstellar HZ may have collected between 4 and 40 Earth
oceans from planetary embryos, 
but main trend to appear was: the more eccentric is the orbit of the binary, the more eccentricity is also
injected into the gas giant's orbit. This in turn leads to
fewer and dryer terrestrial planets. \cite{quintana07} 
emphasize that during the late stages of planetary formation, without the presence of gas 
giant planets, binary stars with periastron $>$ 10 au have a minimal effect on terrestrial planet formation 
within $\sim$2 au of the
primary, whereas binary stars with periastron $\le$ 5 au restrict terrestrial planet formation to within $\sim$1 au of 
the primary star. \cite{quintana06} studied the late 
stages of 
planetary formation in P-type orbits in binary star
systems (with maximum separation equal to 0.4 au and maximum secondary's 
eccentricity equal to 0.8) including Jupiter and
Saturn-like planets. They conclude that the higher the secondary's apoapsis, the 
smaller the number of planets formed and the lower their mass.
The anti-correlation between a system's eccentricity and the number of planets has also been found in a preliminary
interpretation of observation statistics by \cite{limbach14}.
These results indicate that the likelihood of finding habitable planets in such an environment could be small. \\

\noindent
Stochastic simulations proved that planet can also be formed almost dry in the circumprimary HZ of binary star 
systems \citep{haghighipour07}. However, as emphasized by these authors, water delivery in the inner solar system is 
not only due to radial mixing of planetary embryos. Smaller objects can also contribute as shown in \cite{raymond07}. 
Indeed, they highlighted that water delivery from smaller planetesimals is statistically robust and should supply 
terrestrial planets with a significant water source of perhaps three to 10 oceans. In our work, we aim to answer how 
much water can be transported into the HZ via small bodies, remnants from the late phase of the planetary 
formation, providing thus other water sources to embryos. We statistically study the dynamics 
of an asteroid belt in such systems and we treat this problem in a self-consistent manner
as all the gravitational interactions in the system as well as water loss of the planetesimals due to outgassing are
accounted for.
Our main goal is to examine the influence 
of the secondary star on the flux of small bodies from icy regions to the HZ and the efficiency of the
water transport within a short timescale of 10 Myrs.

\section{Initial conditions and dynamical model}
\label{S:initial}

Our study is focused on a primary G-type star with mass equal to one solar 
mass and we investigate the dynamical effect of a secondary of either G, K or M-type with 
physical properties expressed in Table \ref{T:phys}.
\begin{table}
 \begin{center}
  \caption{Physical properties of the secondary star in the binary system.}
  \label{T:phys}
  \begin{tabular}{|c|c|c|c|}
   \hline
   \hline
Stellar-type& M$_{\star}$ [M$_{\odot}$] & L$_{\star}$ [L$_{\odot}$] & 
T$_{\star}$ [K] \\
\hline

 G & 1.0 & 1.0 & 5780 \cr

\hline
 
K & 0.7 & 0.38 & 5200 \cr
\hline
  
 M & 0.4 & 0.08 & 3800\cr

\hline
\hline
  \end{tabular}
 \end{center}
\end{table}
\noindent 
The studied binary star systems encompass relatively tight configurations, i.e.
semi-major axes in a range of a$_{\scriptscriptstyle \text{b}} \in [25:100]$ au.
This parameter has been changed in steps of 25 au in our simulations.
The secondary is on an elliptical coplanar orbit with eccentricities
e$_{\scriptscriptstyle \text{b}} \in [0.1:0.5]$ increased in steps of 0.2. In total 12 
configurations have been investigated for any given stellar type of the companion. In addition, we consider a gas giant 
located at
a$_{\scriptscriptstyle \text{GG}}$ = 5.2 au on a circular coplanar orbit, with a mass equal to Jupiter's mass. \\

\noindent
A disk of planetesimals is modeled as a ring of 10000 asteroids with masses 
similar to main belt objects
in the Solar System and each asteroid was 
assigned an initial water mass fraction (hereafter wmf) of 10\%
\citep{abe00,morbidelli00}. During the late phase of planetary formation, if the inner 
region of the system is mainly 
dominated by large embryos (following a specific mass distribution) with masses 
between Moon to Mars size 
\citep{raymond04,haghighipour07}, debris resulting from collisions of such 
embryos are also present.
 To determine the lower and upper limits for the asteroids' masses, 
we performed independent preliminary simulations with a 3D smooth particle 
hydrodynamics (SPH) code \citep{sch05,maisch13}. First-order
consistency is achieved by a tensorial correction as discussed in 
\citet{schspe07}. It includes self-gravity and models
material strength using the full elasto-plastic continuum mechanics and the 
Grady-Kipp fragmentation model for fracture
and brittle failure \citep{grakip80,benasp94}.
The scenarios involve collisions of rocky basaltic objects with one lunar mass 
at different encounter velocities and 
angles $\alpha\/$. The latter are defined in a way so that $\alpha=0^\circ\/$ 
corresponds to a head-on collision. The
scenarios start with the bodies five diameters apart to let the SPH particle 
distribution settle. This preliminary simulation time-span was 2000\,min.
In an earlier study, most collisions of Moon-sized bodies in the Solar System's 
HZ were found to happen at velocities
$v\la 2\,v_\mathrm{esc}\/$ at arbitrary collision angles \citep{maidvo13}. We 
chose initial conditions in this range
which are in the merging/partial accretion and hit-and-run domain of the 
collision outcome map \citep[cf.][]{agnasp04,leinhardt12,maidvo14b}. Expecting a somewhat higher spread in $v\/$ 
in binary systems we also included a
scenario in the erosion/mutual destruction domain.
Table~\ref{tab:massloss} lists the collision scenario parameters and gives the 
resulting mass loss of the survivors (one
body for merging, two bodies in the erosion and hit-and-run scenarios, 
respectively) after one impact.

\begin{figure*}
    {\includegraphics[width=\textwidth]{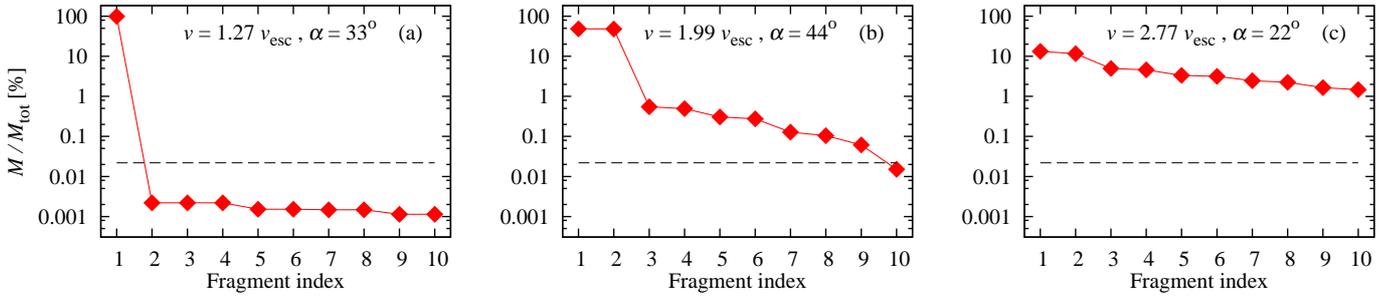}}
\caption{Collision scenarios resulting in a merge (a), a hit-and-run encounter 
(b) and erosion/mutual destruction (c).
The diagrams show the masses of the ten largest fragments at the end of the 
simulation as percentages of the total
system mass $M_\mathrm{tot}=2\/$ lunar masses. The dashed horizontal line 
indicates the limit for significant fragments
(see text).}
\label{fig:scenarios}
\end{figure*}

\begin{table}
\caption{Mass loss in the simulated collisions. Mass loss is defined as the 
quantity not in the surviving bodies (one
for erosion, two for hit-and-run and merging) after one impact.}
\centering
\begin{tabular}{rrlr}
\hline
\multicolumn{3}{c}{Collision scenario} &
\multicolumn{1}{c}{Mass} \\
$v\/$                & $\alpha\/$     \\
$[v_\mathrm{esc}]\/$ & $[^\circ]\/$ & Type  & \multicolumn{1}{c}{[wt-\%]} \\
\hline
1.00 & 20 & merging    &   0.6 \\
1.27 & 33 & merging    &   0.9 \\
1.37 & 42 & hit \& run &   0.4 \\
1.99 & 44 & hit \& run &   3.5 \\
2.77 & 22 & erosion    &   75.2 \\
\hline
\end{tabular}
\label{tab:massloss}
\end{table}
\noindent
The fragment sizes beyond the survivors drop significantly in the hit-and-run 
and merging scenarios
(Fig.~\ref{fig:scenarios}a, b): In the 
mutual destruction case, all of the ten largest fragments possess masses $\ga1\,\%\/$ of the total system mass, which 
is $\sim$ 
Ceres' mass\footnote{Ceres' mass is equal to 
4.73$\times 
10^{\scriptscriptstyle -10} \text{M}_{\scriptscriptstyle \odot}$}, and hundreds are above
the ``significant'' fragment threshold in the sense of 
\citet{maidvo14b}. The smallest fragment consists of one SPH particle (0.001\% of the total 
mass for 100k SPH particles) which corresponds to $\sim$ 0.1\% of Ceres' mass. As increasing the number of SPH 
particles will 
result in even smaller fragments, this mass is an upper limit for the smallest fragment. As this latter will 
contain $\sim$ 0.006\% Earth-oceans units\footnote{1 ocean = $1.5 \times
10^{\scriptscriptstyle 21}$ kg of H$_{\scriptscriptstyle 2}$O}, we chose to neglect the water contribution of 
smaller particles. Our minimum and maximum mass are thus defined according to the fragments' mass after one 
impact. Therefore, members of our ring will have masses
randomly and equally distributed between 0.1\% to one Ceres' mass. The total mass of the ring of 10000 
asteroids amounts to 
M$_{\scriptscriptstyle \text{R}}$ =0.5 M$_{\oplus}$. Thus the quantity of water, in terms of Earth-ocean 
units, available in 
a ring will be 200.\\

\begin{figure}
 \centering{
  \includegraphics[angle=-90, width = \columnwidth]{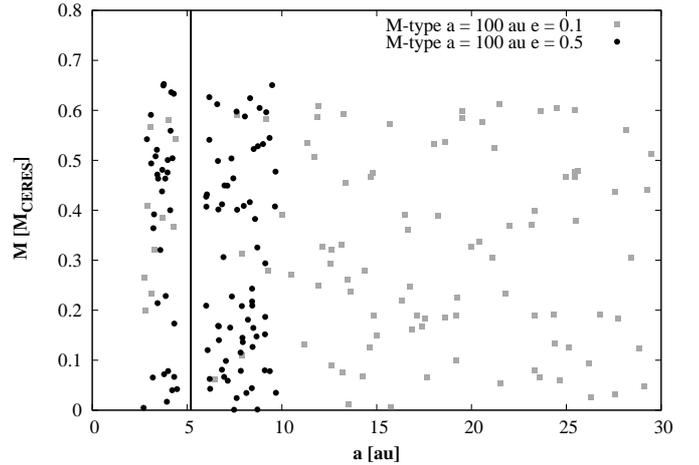}}
\caption{Example of mass distributions in the circumprimary ring as a function of the asteroids' semi-major axis under
the
gravitational influence of
a secondary M star at a$_{\scriptscriptstyle \text{b}}$ = 100 au and 
e$_{\scriptscriptstyle \text{b}}$ = 0.1
(\textcolor{gray}{$\blacksquare$}) and e$_{\scriptscriptstyle \text{b}}$ = 0.5 
(\textcolor{black}{$\bullet$}). The vertical line
refers to the gas giant's position. All particles are initially spaced by several Hill's radii to prevent immediate
dynamical instability.}
\label{F:mass_distrib}
\end{figure}

\noindent
In each system, we defined the size of the disk of planetesimals by the following borders. As we focus on the 
transport 
of icy bodies, the inner border of the disk is set 
to the snow-line position \citep{lecar06,martin12,martin13}, 
border
between icy and rocky planetesimals. Its position changes as the star's luminosity evolves during its birth phases.
As our host star is a G2V type, we considered the inner border of the disk of planetesimals according to observations in 
the 
Solar System, suggesting the snow-line to be at $\sim$ 2.7 au. The outer border of the disk of planetesimals is 
influenced by 
perturbations induced 
by the secondary star. Since we consider initially circular motion for the planetesimals, the stability border depends 
mainly on three parameters:
the mass ratio of the system, a$_{\scriptscriptstyle \text{b}}$ and e$_{\scriptscriptstyle \text{b}}$
\citep{rabl88,holman99,pilat02}. These 
authors showed
that it is possible to link these parameters to derive a
critical semi-major axis a$_{\scriptscriptstyle \text{c}}$ as the maximum 
initial semi-major axis for a particle to survive in the system. In
contrast to \cite{rabl88} and \cite{holman99} who classified unstable orbits via
ejections of test planets from the system, \cite{pilat02} calculated the Fast
Lyapunov Indicator (FLI) for each orbit to distinguish between stable and chaotic motion. This well known chaos 
detection 
method was introduced by \cite{froeschle97}. In case of circular motion of the planets and the planetesimals, it is  
possible to use the study by \cite{holman99} where a$_{\scriptscriptstyle \text{c}}$ and its uncertainty $\Delta
\text{a}_{\scriptscriptstyle \text{c}}$ allows a good determination of the outer border (maximum semi-major 
axis allowed) for
asteroids in the ring  as a$_{\scriptscriptstyle \text{c}}$ - $\Delta \text{a}_{\scriptscriptstyle \text{c}}$, which is
in good agreement with the stability limits derived  from FLI computations \citep{pilat02}. Asteroids of our belt
are
randomly positioned between the inner (the snow-line) and and outer borders (the stability limit a$_{\scriptscriptstyle 
\text{c}}$ - $\Delta
\text{a}_{\scriptscriptstyle \text{c}}$). Figure 
\ref{F:mass_distrib} shows a comparison of
initial
asteroid's mass distribution for a secondary M star at 100 au and
e$_{\scriptscriptstyle \text{b}}$ = 0.1 and 0.5. The position of the gas giant planet is indicated 
by the vertical line. Because of the binary's
tight periapsis in the case of e$_{\scriptscriptstyle \text{b}}$ = 0.5, the asteroid belt is less extended and exhibits
a higher mass density. In order to prevent immediate dynamical instability, all bodies have been
carefully placed so that their initial mutual separation were at least several Hill's radii. \\

\noindent
In our main simulations, as we
assume that the giant planet is already formed so that the remaining gas has been evaporated or coagulated into the 
asteroid belt, 
we do not take any effects related to gas drag into account (therefore no gas driven migration, no eccentricity 
dampening).
The initial eccentricities and
inclinations are randomly chosen 
below 0.01 and 1$^\circ$ respectively.
To avoid strong initial interactions with the gas giant, we assume that the giant planet has cleared a path
in
the disk around its orbit. The width of this gap is $\pm 3\,\text{R}_{\scriptscriptstyle {\text{H},\text{GG}}}$ where
R$_{\scriptscriptstyle 
{\text{H},\text{GG}}}$ is the giant planet's Hill Radius\footnote{However, we did not exclude possible location of
asteroids
inside mean motion resonances (MMRs). Indeed, we assume that the presence of the gas might have kept asteroids on stable
orbit inside MMRs. Then, the MMR perturbations became stronger when the gas vanished.}. Requiring orbital stability of
the gas giant at 5.2 au reduces the number of possible binary configurations.  Therefore,  we excluded the case where
a$_{\scriptscriptstyle \text{c}} - \Delta
\text{a}_{\scriptscriptstyle \text{c}}  \le \text{a}_{\scriptscriptstyle {\text{GG}}} + 
3\text{R}_{\scriptscriptstyle {\text{H},\text{GG}}}$. This results in a total number of 23 binary systems
configurations that were studied in this work and summarized in 
Table \ref{T:conf}.\\
\begin{table}
 \begin{center}
  \caption{Binary configurations studied in this article. Only the secondary 
stellar-type is reported.}
  \label{T:conf}
  \begin{tabular}{|c|c|c|c|c|}
   \hline
   \hline
\backslashbox {e$_{b}$}{a$_{b}$ [AU]}   &  25  & 50 & 75 & 100 \\
  \hline

0.1 & K-M & G-K-M & G-K-M & G-K-M \cr
\hline

0.3 &  & G-K-M & G-K-M & G-K-M \cr
\hline

0.5 &  &  & M & K-M \cr

\hline
\hline
  \end{tabular}
 \end{center}
\end{table}

\noindent
We limited our study to 10 Myrs integration time. This is of course much shorter than the timescale of 
terrestrial planetary formation ($\sim$
100 Myrs) but as we make a comparison of many different binary star systems, we had to restrict this study to a
shorter time. However, we will select from this study the most interesting systems for which a statistic over 100 Myrs 
will be made. We numerically
integrated our systems using the \textit{nine} package 
\citep{eggl10} and only
gravitational perturbations were taken into account. The numerical integrator used for 
the computations is based on Lie-series (see e.g. \cite{hanslmeier84} and more recently \cite{bancelin12}). For a 
given configuration in Table \ref{T:conf}, as our planetesimals do not interact with each other, the disk was divided 
into 100 subrings and separately integrated.

\section{The Habitable Zone borders}\label{S:HZ}

As we study the flux of water-rich asteroids into the HZ, we need to know the position of this area. The 
definition, modeling and computation of
the "classical HZ" (hereafter CHZ) is given in \cite{kasting88}, \cite{kasting91}, \cite{kasting93} and
\cite{kopparapu13}. All these studies are based on 
a 1D cloud-free climate model, where the inner edge of the HZ is computed by increasing the surface temperature and the
outer
edge, by increasing the CO$_{\scriptscriptstyle 2}$ partial pressure (maintaining a constant surface temperature at 273
K). The corresponding stellar flux, needed to maintain the surface temperature, is then derived. Thus, these authors
were able to express the CHZ borders as a simple fit function containing the stellar luminosity and temperature.
Recently, \cite{kopparapu14} 
investigated the dependence of the HZ
borders on the planetary masses and derived new coefficients for the 
computations of the effective solar
flux. For a G star, the inner edge of the HZ corresponding to the runaway 
greenhouse limit is
0.950 au and the
outer edge of the HZ corresponding to the maximum greenhouse is 1.676 au. The updated inner edge value is closer to the
Sun
than the one found by \cite{kopparapu13} because they used inputs from a 3D model by \cite{leconte13}. However, the
exact value for the inner CHZ border depends on many assumptions and is still under discussion in current
literature, \citep[e.g.][]{wolf14}. \\

\begin{figure}
  \centering{
 \includegraphics[angle=-90,width=\columnwidth]{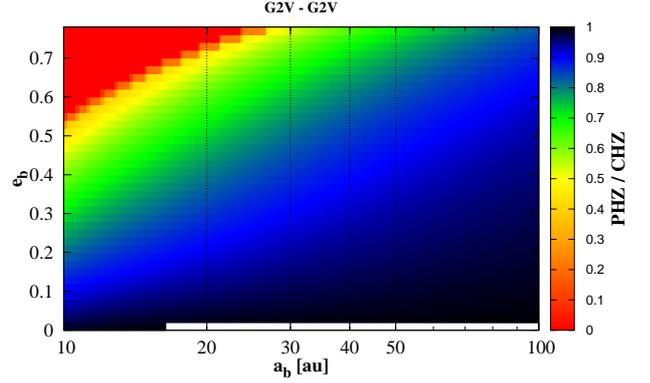}}
  \caption{PHZ/CHZ ratio as a function of the binary separation and the 
secondary's eccentricity. The red color indicates
that the planet will always lie outside the PHZ borders. 
Only a
graph for a secondary G star is represented as it is quite similar for a secondary K and 
M-types.}
  \label{F:PHZ}
\end{figure}

\noindent
 When 
including a binary companion in the
system, the additional gravitational interaction and radiation can shrink the CHZ borders, as shown in
\cite{eggl12} and \cite{kaltenegger13}. As a matter of fact, the smaller the periapsis of the binary, 
the more important the insolation of the primary, as the periapsis distance of the planet with respect to its host star
can change significantly. In order to account for this effect, \cite{eggl12} introduced the so-called
Permanently
Habitable Zone (hereafter PHZ). The PHZ contains information on the planet's perturbed orbit. It serves as a mean to
distinguish 
areas where a planet always receives the correct amount of insolation to remain habitable from regions 
where insolation conditions for habitability are only fulfilled in an average\footnote{The planet's orbit is eccentric
enough could leave the HZ from time to time.} sense. Figure \ref{F:PHZ} shows the ratio PHZ/CHZ as a 
function of the binary's orbital elements (a$_{\scriptscriptstyle \text{b}}$,e$_{\scriptscriptstyle \text{b}}$) for a
G2V-G2V binary star system. The blue and black colors cover a large region where no or only minor differences between  
PHZ and CHZ were found. This means that the additional radiation from the second star is not
enough to drastically cause a change of insolation at the surface of a planet. This is due to the fact that the
secondary is too far or its periapsis too far away. For higher eccentricities, where the secondary's orbit approaches 
closer to the host star, the difference is increased (green color).
The red color
refers to regions where the PHZ vanishes either due to excessive insolation or due to orbital instability.
It is clear that the truncation of the PHZ
increases for large values of the periapsis of the binary companion. This is due to the fact that the stellar
gravitational perturbations acting on the planet will
cause a significant change of the planet's periapsis, which in turn influences the insolation at the planet's surface.
A similar behavior is observed for K and M class
secondaries as their mass does not significantly influence the PHZ/CHZ ratio.
\noindent
For our binary configurations, the difference between CHZ and PHZ is only of the order of 10\% as shown in Table 
\ref{T:PHZ}. As these
values do not vary strongly with the secondary's mass, we did not report all 23 
configurations. \\
\begin{table*}
 \begin{center}
  \caption{PHZ borders given in au as a function of the binary orbital characteristics (a$_{\scriptscriptstyle
\text{b}}$, e$_{\scriptscriptstyle \text{b}}$) used
in our simulations. Compared to the CHZ borders (inner border is 0.950 au and outer border is 1.676 au), the PHZ
intervals are smaller due to the additional radiation from the second star and eccentricity injected in the planet's
orbit.}
  \label{T:PHZ}
  \begin{tabular}{|c|c|c|c|c|}
   \hline
   \hline
\backslashbox
{e$_{\scriptscriptstyle \text{b}}$} {a$_{\scriptscriptstyle \text{b}}$ [au]}   &  25  & 50 & 75 & 100 \\
  \hline

  0.1 &[0.959:1.654]  & [0.955:1.664] & [0.954:1.668] 
 & [0.953:1.670]\cr

\hline

  0.3 & & [0.964:1.639]  &[0.959:1.651]  & [0.957:1.657] 
\cr
      
\hline

  0.5 &  & &[0.968:1.627] & [0.964:1.638]\cr

\hline
\hline
  \end{tabular}
 \end{center}
\end{table*}

\section{Asteroid flux and water transport to the HZ}\label{S:numerical}

\subsection{Statistics on the disk dynamics}

\begin{figure*}
  \centering{
 \includegraphics[angle=-90,width=0.7\textwidth]{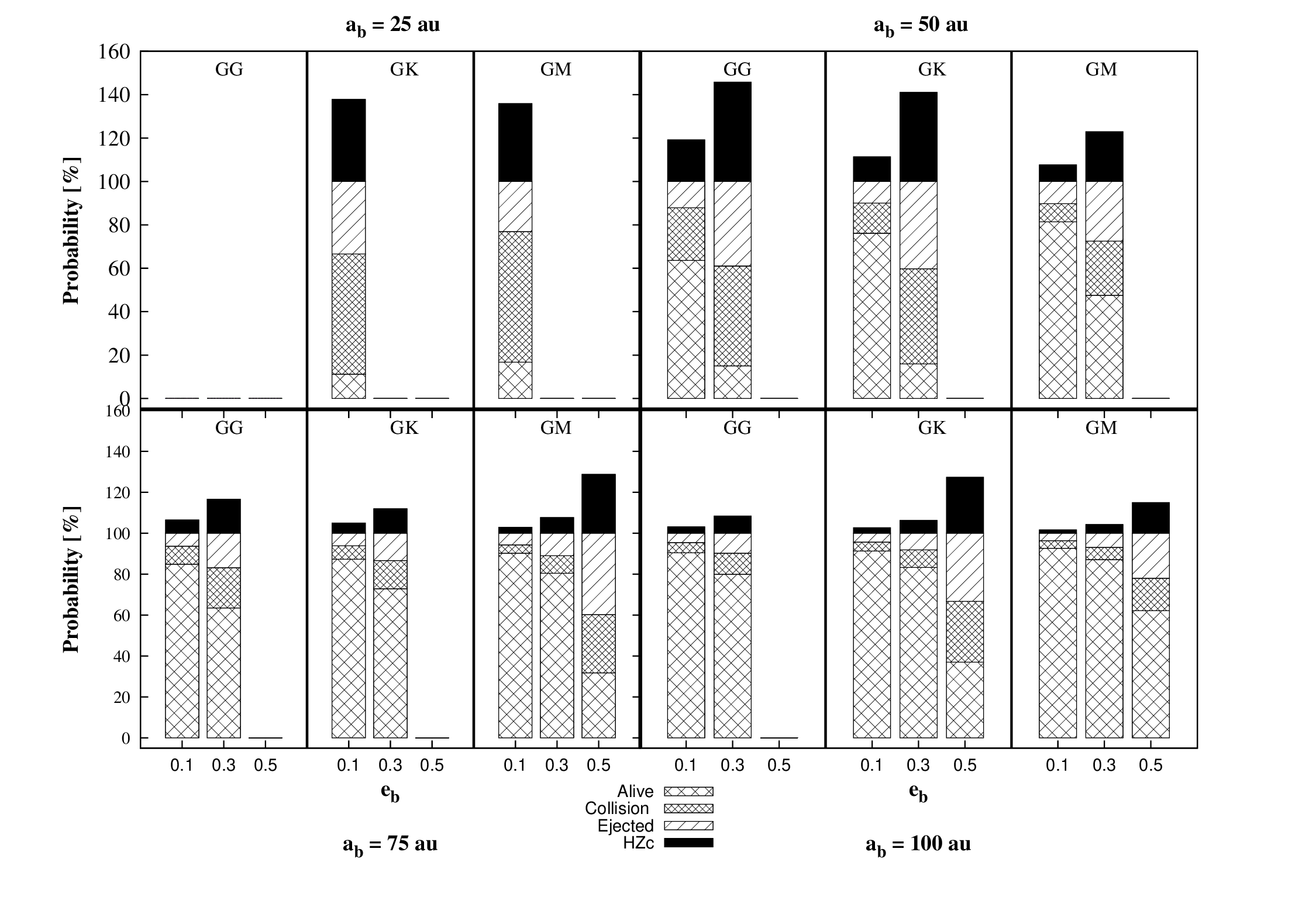}}
  \caption{Statistics on the disk of planetesimals dynamics. Each histogram shows the 
evolution (expressed in probability) of the
asteroids in the ring within 10 Myr of integration. They can still be present in the system 
("alive"), become a HZc, collide with
the stars or the gas giant,  or be ejected out of the system. The closer the secondary 
and the higher its mass, the higher the
probability to empty the asteroids ring.}
    \label{F:impactor}
\end{figure*}

During the simulation, each particle is tracked until the end of the
integration time in order to assess the following numbers:

\begin{itemize}
 \item asteroids crossing the HZ. They will be referred to as Habitable Zone crossers\footnote{A HZc can cross several
times the HZ 
before leaving the system or colliding
with the stars or the planets} (hereafter HZc). As we assume a two dimensional HZ, an asteroid will be considered as a
HZc if the intersection point between its orbit and the HZ plane lies within the HZ borders.
 \item asteroids leaving the system when their semi-major axis$\ge$
500 au
 \item asteroids colliding with the gas giant or the stars
 \item asteroids still alive in the belt after 10 Myr
\end{itemize}
\noindent
 Figure \ref{F:impactor} shows the resulting statistics on the asteroids' dynamics. It represents four
panels corresponding to the secondary's semi-major axis investigated. Each histogram shows the dynamical outcome 
of our asteroids
expressed in terms of probability, as a function of the secondary's eccentricity. Below 100\%, the percentage of 
asteroids that are still present in the belt ("alive"), that were ejected or collided with the
stars or the gas giant is shown. The black area of each histogram above 100\% indicates 
the probability for asteroids
to
enter the HZ. Such asteroids, crossing the HZ, are called HZc. The 
perturbations due to both the gas giant and the
binary
companion can cause an increase in eccentricity of the asteroids within the planetesimal disk which of course depends 
on 
the  binary configuration. A comparison of the different histograms indicates that the key parameter is the periapse of 
the binary system which is defined by the eccentricity of the binary. Of course not only the asteroids are perturbed by 
the secondary, the gas giant at 5.2 au is also perturbed and its initially circular motion will change to an elliptic 
one. This behavior is highlighted by the 
increasing value of the probability
to become a HZc if the secondary's periapsis distance decreases and if its mass increases. Indeed, for a given value of 
a$_{\scriptscriptstyle b}$ (for instance 50 au), one can see that
this probability is at least doubled when e$_{\scriptscriptstyle b}$ increases. As a consequence, the
asteroid belt will be depopulated because of dynamically induced ejections 
as well as collisions with the giant planet
and the stars. Because the rate of colliding and ejected asteroids is increased, a ring will be depopulated faster when
e$_{\scriptscriptstyle b}$ becomes larger. Therefore, the statistics in Fig. \ref{F:impactor} shows that the 
probability for a member of the asteroid
ring to stay in the system after 10 Myrs will
decrease with the periapsis distance and the secondary star's mass. Finally, these results can also answer the question
of the
presence of an asteroid belt in such systems. Indeed, if we assume that the gas could protect small bodies from secular
resonances or mean motion resonances (MMRs), it is highly unlikely that they survived in binaries with small 
periapsis separation, after the gas has dissipated.\\

\subsection{Timescale statistics}\label{SS:timescale}

\begin{figure*}
  \centering{
 \includegraphics[angle=-90, width=0.7\textwidth]{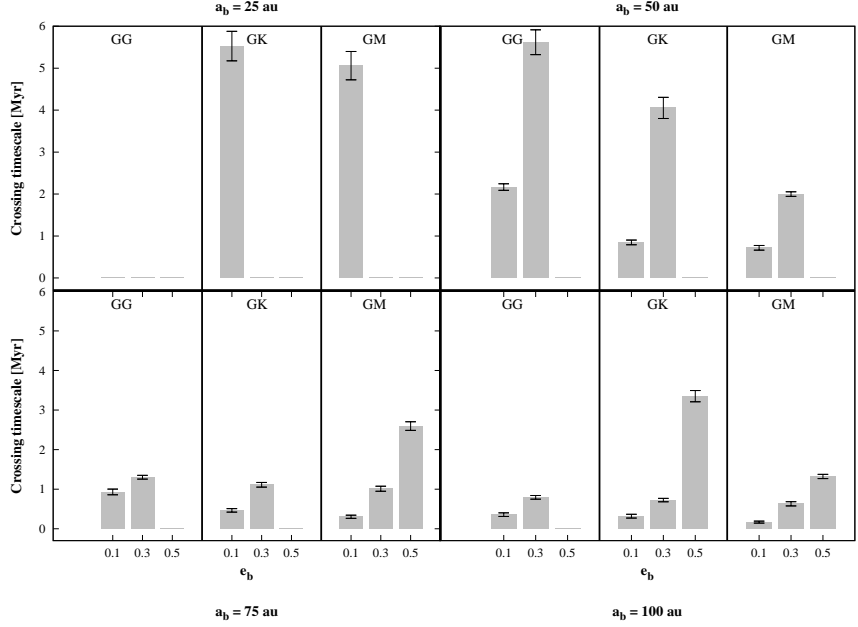}}
  \caption{Crossing timescale expressed in Myr. It corresponds to the last crossing inside the HZ. Up
to 6 Myr for the most perturbed systems but also with the lowest
number of asteroids still present in the system within 10 Myr of 
integration time. For the other systems, the crossing timescale is much
lower because their belt still contain a huge number of "alive" asteroids. 
Indeed, the dynamics of the system needs more time to empty the 
asteroid belt population.}
    \label{F:timescale}
\end{figure*}

\noindent
Depending on the periapsis distance of the secondary, the disk of planetesimals can be 
perturbed more or less rapidly. As a
matter of fact, asteroids will suffer from the gravitational perturbations of the 
secondary star and the gas giant, and their eccentricity may increase quickly. 
\begin{figure}
  \centering{
 \includegraphics[angle=-90, width=\columnwidth]{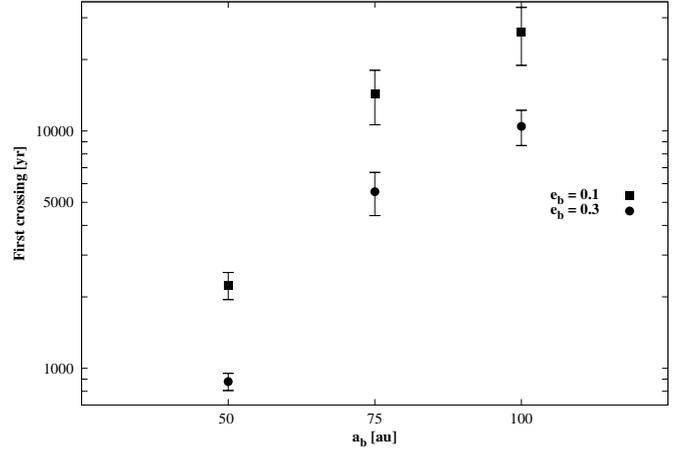}}
  \caption{Median time for an asteroid to become a HZc for the case of a secondary G-type with 
e$_{\scriptscriptstyle \text{b}}$ = 0.1 ($\blacksquare$) and e$_{\scriptscriptstyle \text{b}}$ = 0.3 ($\bullet$). 
This corresponds to the time when an asteroid crosses the HZ for the first time. The statistics is made over the 10000 
asteroids and the 1$\sigma$ value is represented by error bars.}
    \label{F:time}
\end{figure}
Figure 
\ref{F:time} shows statistical results of
the average time needed by an asteroid to become a HZc, i.e. the time it takes to reach the HZ.  This corresponds to
the time spans until the first asteroid enters the HZ. The median value and
its absolute deviation (error bars) are presented for a set of 10000 asteroids for the case of a secondary 
G-type with e$_{\scriptscriptstyle \text{b}}$ = 0.1 ($\blacksquare$) and e$_{\scriptscriptstyle \text{b}}$ = 0.3 
($\bullet$). This confirms a strong correlation between the periapsis distance and the time of first
crossing. Figure \ref{F:time} clearly shows that the average time varies from a few centuries to tens 
of thousands of
years. The
closer and more massive the secondary star, the sooner asteroids can reach the HZ. In contrast, the
crossing timescale will become larger as the number of HZc increases as shown in 
 Fig. \ref{F:timescale}. This parameter corresponds to the bombardment timescale, within 10 Myr of integration 
time, and is derived regarding the last crossing inside the HZ but regardless the asteroid's water content. Indeed, 
once an
asteroid becomes a HZc, it can cross the HZ several times as long as its orbit is stable and
until it is ejected out of the system or collides with the giant planet or the stars. 
However, water-rich asteroids could be dry before this corresponding timescale, revealing that water transport could 
occur on a very short time (compared to planetary formation timescale). Besides, one can notice from Fig. 
\ref{F:timescale} that most of the systems, with large a$_{\scriptscriptstyle b}$ and low e$_{\scriptscriptstyle b}$, 
having low crossing timescales are also those where most asteroids remained in the after 10 Myrs as shown in Fig. 
\ref{F:impactor}. This suggests that asteroid flux to the HZ could occur in several steps as some asteroids need more 
time to 
drastically increase their eccentricity in order to be moved to lower orbits and reach the HZ. This reveals also 
that water sources can still be available in the ring, provided that asteroids still have icy water on their surface. 
This 
study would of course require 
longer integration time $>$ 10 Myrs. Nevertheless, we have a clear guess on the efficiency of binary stars systems to 
transport asteroids from beyond the snow-line to the HZ on a short timescale. 

\subsection{Water transport statistics}

Each HZc entering the HZ will bring a certain amount of water. However, regarding the integration
timescale, the water content of asteroids may vary with time. Indeed, increased eccentricities can cause asteroids to
approach close to the stars. This in turn would lead to a mass variation mainly due to a loss of their water content. 

\subsubsection{The water mass loss process}

To quantitatively assess the water content of asteroids, we followed their water mass fraction evolution 
throughout the simulations, including 
mass loss process, until they enter the HZ. The main mechanisms that can induce a relevant mass loss for 
active (comet-like) or inactive
asteroids are ice
sublimation and impact ejection \citep{jewitt12}. The latter process occurs when smaller asteroids impact
larger ones\footnote{Such impacts do not necessarily lead to a complete destruction or break-up of asteroids}. These
impacts can be highly erosive due to characteristic speeds $\sim$
5 km.s$^{\scriptscriptstyle -1}$ \citep{bottke94} and the ejected amount mass can be significant. However, typical
impact events in the
main-belt have an impact probability $\sim$ $3.0\times 10^{-8}$ km$^{\scriptscriptstyle {-2}}$.yr$^{\scriptscriptstyle
{-1}}$ \citep{farinella92}. Yet, according to \cite{bottke05}, the timescale for an impact event to happen in our
sample of asteroids belt (with radius from tens to hundreds of kilometers) is much longer that our integration
time. Thus, we neglected this process in our
study. The
only mass loss process we consider is due to ice sublimation when the asteroid comes 
relatively close to the star. This process can be
stepped up in double star systems, especially when the companion is on an eccentric orbit. 
Even if the secondary is not particularly close, asteroids' eccentricity can be pumped and they can receive a
large amount of insolation. 
Therefore, the water mass loss rate
$\dot{m}$ was computed accounting for the radiation of both stars. The estimation of 
$\dot{m}$ can be derived by solving a
balance energy equation between the net incoming stellar flux, the ice 
sublimation and the thermal re-radiation\footnote{assuming the asteroid to have black body properties}, as expressed in
Eq. (\ref{E:energy})
\begin{equation}
\label{E:energy}
 \sum_{i=1}^2 \frac{\text{F}_{\star,\text{i}}(1-\text{A}_{\scriptscriptstyle \text{i}})}{\text{R}^{\scriptscriptstyle
2}_{\scriptscriptstyle 
\text{i}} [\text{au}]}\cos{\theta}_{\scriptscriptstyle \text{i}} = \epsilon \sigma
\text{T}^{\scriptscriptstyle 4} + \text{L(T)}\,\dot{m}\text{(T)} 
\end{equation}
\noindent
where
\begin{itemize}
 \item $\text{F}_{\star}$ is the stellar constant and is computed as 
$\displaystyle{\text{F}_{\star} =
\text{F}_{\odot}\,{\text{L}_{\star}}}$ (F$_{\odot}$ $\sim$ 1360
W.m$^{\scriptscriptstyle{-2}}$ is the solar constant).

\item $A$, the bound albedo, product of the geometric albedo and the phase 
integral, defines the fraction of the
total incident stellar radiation reflected by an object back to space. 
Asteroids can have \textit{A}$ \simeq$ 0.5  but most of
them have relatively low albedo \citep{shestopalov11}. Ice material is known to be a good 
radiation reflector. In order to maximize $\dot{m}$, we will
follow the approach of \cite{jewitt12} considering our objects with dirty ice 
material with low albedo because clean
ice sublimates too slowly at main-belt distances. Therefore, we used an averaged albedo $A_{\scriptscriptstyle 1} =
A_{\scriptscriptstyle 2} = \bar{A} =  0.05$.
\item $R$ is the distance to the star (primary or secondary) expressed in au
\item $\theta$ is the angle between the incident light and the normal to the 
the asteroid's surface. According to
\cite{jewitt12}, $\dot{m}$ is weaker for an isothermal surface than at the sub-solar point 
of a non-rotating body. Again, to maximize $\dot{m}$, we consider the sub-solar case with $\theta
= 0^\circ$.
\item $\epsilon$ is the emissivity of the surface, $\epsilon \sim 0.9$
\item $\sigma = 5.67\times 10^{\scriptscriptstyle {-8}}$ W.m$^{-2}$.K$^{-4}$ is 
the Boltzmann constant
\item T is the equilibrium temperature at the surface expressed in K
\item L is the latent heat of sublimation in J.kg$^{\scriptscriptstyle -1}$
\item $\dot{m}$ is the surface mass loss rate in kg.m$^{\scriptscriptstyle 
-2}$.s$^{\scriptscriptstyle -1}$
\end{itemize}
\noindent
$\dot{m}$ can be obtained by computing the temperature T solving Eq. 
(\ref{E:energy}). To this purpose, we expressed all
the variable parameters as a function of T. According to \cite{delsemme71}, $\dot{m}$ can be written as:
\begin{equation}
\label{E:mass}
 \dot{m} =\text{P}_{\scriptscriptstyle \text{S}}\,\sqrt{\frac{\mu}{2\,\pi \text{kT}}}
\end{equation}
\noindent
where $\mu = 18$ g.mol$^{\scriptscriptstyle -1}$ is the water molar mass and 
$\text{P}_{\scriptscriptstyle \text{S}}$ is the
saturation pressure and defined by the empirical formula: 
\begin{equation}
\label{E:pression}
 \log{\text{P}_{\scriptscriptstyle \text{S}}} =  4.07023-\frac{2484.986}{\text{T}} + 3.56654\log(\text{T}) 
- 0.00320981\,\text{T}
\end{equation}
\noindent
Finally, the latent heat is expressed as:
\begin{equation}
\label{E:latent}
 \text{L (T)} = 2834.1 -0.29\,(\text{T}-273.15)-0.004\,(\text{T}-273.15) 
~~\text{J.g$^{\scriptscriptstyle -1}$}
\end{equation}

\subsubsection{The water mass fraction of incoming HZc}\label{SS:water}

The value of $\dot{m}$ is constantly updated during the simulations and the accumulated surface mass loss $dm$ reads:
\begin{equation}
\label{E:accumulated}
 dm =  \sum  \dot{m}\,\Delta t ~~~~~~\mbox{kg.m$^{\scriptscriptstyle -2}$}
\end{equation}
\noindent
where $\Delta t$ represents the time elapsed since the beginning of the 
integration. We then compute the sublimating area $4\pi\,r^2$ with $r$ the radius of the HZc. If we consider the
following
density for water ice shell and a basalt core, respectively $\rho_{\scriptscriptstyle \text{i}} = 900$
kg.m$^{\scriptscriptstyle -3}$ $\rho_{\scriptscriptstyle \text{c}} = 3000$ kg.m$^{\scriptscriptstyle
-3}$, then for a given wmf, the mean density of an asteroid is 
\begin{equation*}
\bar{\rho} = (1-\text{wmf})\,\rho_{\scriptscriptstyle \text{c}} +
\text{wmf}\times\rho_{\scriptscriptstyle \text{i}}
\end{equation*}
Therefore
\begin{equation*}
 \displaystyle \text{r} = \left (
\frac{3\,\text{m}}{4\,\pi\bar{\rho}}\right)^{1/3}
\end{equation*}
with $m$ the mass of the asteroid. Thus we can derive the total water mass loss in kg
 $\Delta\,m
= dm\times 4\pi\,r^{\scriptscriptstyle 2}$. \\

\begin{figure*}
  \centering{
 \includegraphics[angle=-90, width=0.7\textwidth]{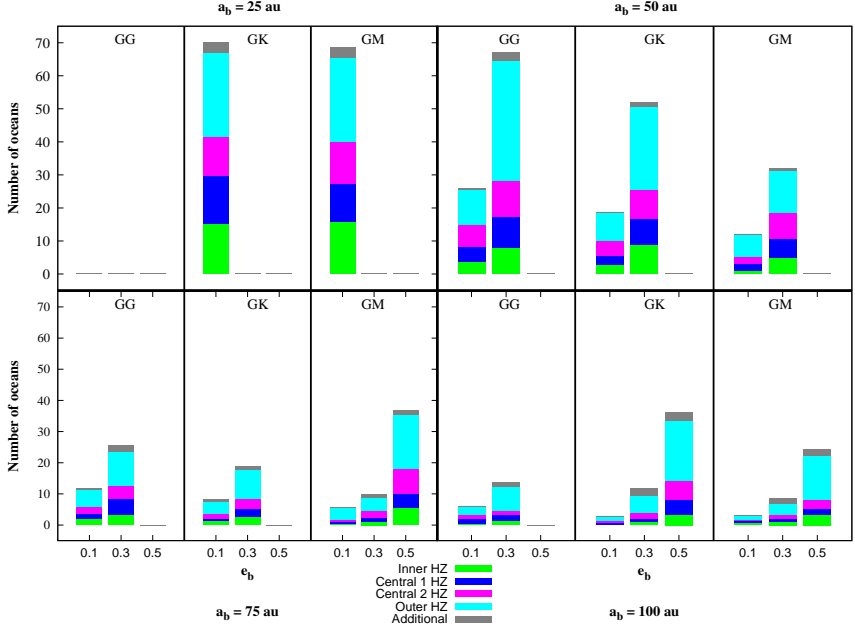}}
  \caption{Total number of oceans transported by all the HZc when 
they first cross the HZ. The color
code is related to four equally spaced rings of the HZ: inner, central 1 and 2, outer HZ (see text). We also 
indicated the equivalent number of
additional ocean crossing the truncated CHZ interval when the dynamics of the 
secondary star is not taken into account
for the computation of the HZ borders.}
    \label{F:ocean}
\end{figure*}

\noindent
Whenever an asteroid becomes a HZc i.e. when it first enters the HZ, we
suppose that its current water content will be delivered. This is the best case scenario as the total water
content of the HZc corresponds to the maximum amount of water available in the HZ. In reality, however, only a small
fraction of water would be accreted onto a planet (or planetary system) inside the HZ. Indeed, both the real position of
the planet on its orbit and the impact velocity of these HZc are key parameters in collisional material transport as
pointed out by \cite{thebault06} and \cite{leinhardt12}. A fully self consistent modeling of the water delivery at
impact lies beyond the scope of the current work, however. Hence, we aim to study the amount of water transported
into the HZ rather than the amount of water accreted by a planet therein. Figure \ref{F:ocean} shows the
total amount of water that was transported into the HZ (expressed in terrestrial oceans units) as a function of the
binary system characteristics. Each histogram represents the total amount of oceans ending in four partitions of the HZ.
They correspond to equally spaced rings using the values obtained in Table \ref{T:PHZ}. Therefore we define the inner
HZ border (closest ring to the PHZ inner border value), Central 1 and 2 (intermediate borders) and Outer HZ (closest
ring to the PHZ outer border value). We also illustrated on this figure, the additional
quantity of water transported between the inner edge $A$ $[CHZ(A);PHZ(A)]$ and the outer edge $B$ $[PHZ(B);CHZ(B)]$ 
of the HZ (gray color and refereed as "additional"). \\


\noindent
The main results suggested by this figure are:

\begin{itemize}
\item[a)]one can see that a maximum of 70 oceans can be transported into the CHZ. Note that 70 oceans represents 35\% of
the total number of oceans contained in the 10000 asteroids
population\footnote{200 oceans contained in this population}. Variances are large, however. One of the reasons for
the large spread in the amount of transported water is the short simulation time. This introduced a bias towards
systems where the water transport is fast. In other words, the seemingly small amount of transported water in some of
the presented systems does not imply that their HZs have to remain dry. It merely highlights the fact that the
delivery process would require a longer time as many asteroids are still present in the system after 10 Myrs 
of integration time (see Figure \ref{F:impactor}). However, these results show how fast and efficient are some systems
to transport water.\\

\item[b)] the difference between the water transported in 
CHZ and PHZ (gray color) is not that
significant compared to the total number of incoming oceans. Indeed, the difference does not exceed $\sim$3
oceans. One should note that we disregarded large orbital variations of any embryos located in the HZ caused by the
perturbed motion of the gas giant because of the presence of the secondary star. This in turn can perturb the motion on 
any planets in the HZ\footnote{when including a gas giant in a binary star systems, its additional perturbation will 
increase the eccentricity on any planets located in the HZ}.
However, according to \cite{williams02}, an average insolation -- covering almost the entire CHZ -- is sufficient to
retain liquid water on an Earth-twin surface and secure the habitability of planets. \\

\item [c)] any binary star systems is efficient enough to produce a flux of asteroid within the whole HZ and thus to
make water sources available for embryos lying there. However, we can see that statistically, most of the water ends in
the outer HZ. This is due to the fact that its surface area is much more wider than the other rings and the
probability for an asteroid to fell in the outer HZ is higher. One can notice that the quantity of water 
brought into the outer HZ versus the other cells is quite balance. Indeed, regarding the definition of a HZc, we 
expect asteroids on inclined orbits not to necessarily cross the outer HZ. Besides, dynamical 
studies suggest that secular perturbations can lie inside the HZ or beyond the snow-line, depending on the binary 
systems' characteristics \citep{bancelin15,baszo15,pilat15}. In addition, depending on the secondary's periapsis, the 
secondary will shorten the lifetime of particles inside MMRs \citep{bancelin15}. Therefore, for asteroids initially 
orbiting inside MMRs and/or secular perturbations, the dynamical evolution can be violent and their orbit crossing the 
HZ will not necessarily be from the outer HZ to the inner HZ.

\end{itemize}

\subsection{Influence of the initial water content}

Let us now compare the water transport efficiency when the initial wmf is
not equally distributed throughout the asteroid belt. The equal water distribution model will be called
WMF$_{\scriptscriptstyle
\text{A}}$. Observations in the main-belt
\citep{abe00,demeo14} suggest that a
gradient exists in the
chemical composition of the asteroids. Besides, we expect distant asteroids, up to the Kuiper-belt distance, to have a
higher wmf than asteroids in the main-belt. Thus, we modeled -- we called this model WMF$_{\scriptscriptstyle
\text{B}}$ -- the wmf distribution as a linear function of the
distance to the central star, fulfilling the following boundary conditions: the upper limit is fixed at 20\% of water
for asteroids at 30 au\footnote{As shown in Figure \ref{F:mass_distrib}, our objects' semi-major axis does not go beyond
30 au}. To find the lower limit for asteroids at 2.7 au, we follow \cite{raymond04} and \cite{haghighipour07} assuming a
wmf
$\sim$ 5\%. Under these assumptions, a belt in tight binary systems will
have a lower water mass ratio than a belt under the influence of a secondary on a low eccentric orbit. 
Indeed, the ratio between model WMF$_{\scriptscriptstyle \text{B}}$ and WMF$_{\scriptscriptstyle \text{A}}$ gives lower
limits of 0.62, 0.58 and 0.63 respectively for G, K and M secondary stars. When the
secondary's periapsis increases, we get upper limits of 1.07, 1.18 and 1.27 respectively. However, the quantity of
oceans transported to the HZ using WMF$_{\scriptscriptstyle \text{B}}$ does not exceed 2/3 of the amount of
transported water when using
WMF$_{\scriptscriptstyle \text{A}}$, even if the belt initially
contains more water. This shows that:

\begin{itemize}

\item [a)] our results are robust
\item [b)] asteroids closer to the snow-line are more likely
to become HZc. As a matter of fact, according to model WMF$_{\scriptscriptstyle \text{B}}$
asteroids will contain 10 \% of water if they are located below 11 au\footnote{provided that the critical semi-major
axis
criteria allows asteroids
to be located beyond this distance}. If asteroids becoming HZc were initially beyond this distance, we would have had
an equivalent or higher number of transported oceans.

\end{itemize}

\section{Comparison with a single G star system}

\begin{figure}
  \centering{
 \includegraphics[angle=-90, width=0.45\textwidth]{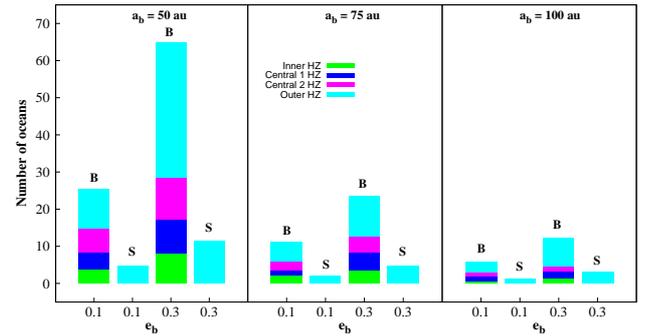}}
  \caption{Comparison of the water transport in single star (S) and binary star (B) systems. We restricted this study
to binaries with two G-type stars. The various panels correspond to different semi-major axis of
the secondary: a$_{\scriptscriptstyle \text{b}}$ = 50 -- 75 -- 100 au. The color code refers to different sub-rings of
the HZ (see text). Note that, the two results for each panel account 
for different initial asteroid belt distributions (depending on 
e$_{\scriptscriptstyle \text{b}}$), taken as the same for (S) and (B).} 
    \label{F:oceanSS}
\end{figure}

In this section, we will compare the water transport efficiency between binary and single star systems. To this
purpose, we considered the same initial conditions for the gas giant and the asteroid
belt distribution, except that we only have a single G
star in our dynamical system. As the comparisons are made for the same asteroids' initial conditions, we have 
to 
consider, for the single star system, the same size for the disk, given by the binaries' characteristics. This 
is why, in Fig. \ref{F:oceanSS}, results for the reference single star case are different depending on the values of 
the binary's eccentricity and separation. 
\begin{figure}[!h]
  \centering{
 \includegraphics[angle=-90, width=\columnwidth]{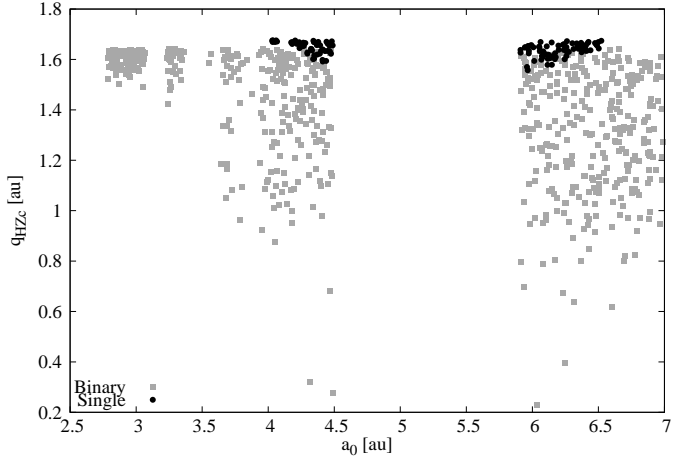}}
  \caption{Initial semi-major axis a$_{\scriptscriptstyle 0}$ and final periapsis distance q$_{\scriptscriptstyle
{\text{HZc}}}$ of the
incoming HZc
for a single (\textcolor{black}{$\bullet$}) and binary (\textcolor{gray}{$\blacksquare$}) star system.}
    \label{F:distrib_a}
\end{figure}

For 
computational 
reasons, our comparison is limited to a G2V-G2V-type binary. Besides, 
regarding Fig. \ref{F:timescale}, there is no more crossings in the HZ after 5 -- 
6 Myr. Therefore, the number of oceans brought to the HZ will not vary after 
this time. Thus, each system was
compared at equivalent integration times
($\le$ 5 Myr). Our results show that an asteroid that was initially in the ring will need 2
-- 20 times longer to reach the HZ in a single star system. That is, because the asteroid belt is not perturbed
strongly enough by the gas giant to produce a large asteroids flux towards the HZ as shown in Fig. \ref{F:distrib_a}:
the
presence of the secondary star can put asteroids with initial semi-major axis a$_{\scriptscriptstyle 0}$,  on high
eccentric
orbits, i.e. very low periapsis distance q$_{\scriptscriptstyle {\text{HZc}}}$. This will result in a faster depletion
of
the belt and a shorter bombardment duration, compared to a single star 
system. 

Consequently, the probability for an asteroid to cross the HZ will be $\sim$
4 -- 6 times higher in a binary
star system. Note that the number of scattered incoming asteroids mainly increases because the secondary star
perturbs the giant planet's orbit. As the flux of asteroids is less efficient in a single star system, the
amount of
water brought to the HZ is smaller. In fact, the water transport is 4 -- 5 times less efficient without a second star.
\noindent
Finally, Fig. \ref{F:oceanSS} compares the amount of delivered water in various systems. Histograms marked with the
letter \textit{B}
refers to binary star systems, those with letter \textit{S} to single star systems. The color code indicates the amount
of water that ended up in four equally spaced sub-rings (Inner, Central 1, Central 2, Outer) of the
corresponding PHZ (see Table \ref{T:PHZ}). For a
single star system, each ring of the CHZ is computed using the
inner edge value 0.950 au and outer edge value 1.676 au. It is not 
surprising that the outer HZ is the most
crossed ring. Indeed, its area is much larger than the other rings. This figure also
highlights the fact
that in such single star systems, hosting only one giant planet, basically all the water is transported in the outer HZ 
because the perturbation is not strong enough to drastically increase the eccentricity of any asteroid in the
belt. In any case, these results show the efficiency of a binary star to transport water in the entire HZ over a shorter
timescale compared to a single star system.

\section{Summary and conclusion}
\label{S:conclusion}

In this work, we investigated the influence of a secondary star on the flux of asteroids to the Habitable Zone (HZ)
over 10 Myrs of integration time. We estimated the quantity of water brought by asteroids located beyond the snow-line 
into the HZ of various double stars configuration (separation, eccentricity and mass of the secondary). An 
overlap of perturbations from the secondary and the giant planet in the primordial asteroid belt causes rapid and 
violent changes in the asteroids' orbits. This leads to asteroids crossing the HZ soon after the gas in the system has 
dissipated and the gravitational dynamics become dominant. Our results point out that binary systems are 
more efficient for water transport into the HZ, compared to a single star system. Not only asteroids flux is  4 -- 6 
times higher when a secondary star is present, but also the amount of transported oceans to the HZ can be 4 -- 5 times 
more important, providing other water sources to embryos, in the whole HZ, in the late phase of planetary 
formation. 
Our results stress the fact that some systems can complete their water transport in a short time ($\sim$ 6 Myr), in 
contrast to single star system. Finally, our study can give a clear guess on the dynamics and the stability of objects 
moving under the perturbations of a binary star system and a gas giant. Indeed, as the depletion of an asteroid belt in 
binaries with small periapsis separations is faster, only a few small bodies will remain member of this belt. Thus, it 
would be unlikely to observe such an asteroid belt in such systems. \\

\noindent
It is clear that dynamics in single and binary systems are completely different as the presence of the 
secondary together
with a gas giant would directly have an impact on the dynamics of an asteroid belt but also on any planets or embryos  
located in
the HZ. Indeed, similarly to the work of \cite{pilat08}, secular perturbations and MMRs' intensity will depend 
on the secondary's orbital parameters and mass, which will be the purpose of a future study. 

\begin{acknowledgements}

DB, EPL, TM and RD acknowledge the support of the FWF NFN project: "Pathways to Habitability" and related subprojects
S11608-N16 "Binary Star Systems and Habitability"  and S11603 "Water transport". DB and EPL acknowledge also the Vienna
Scientific Cluster (VSC project 70320) for computational resources. SE has been supported by the European Union
Seventh Framework Program (FP7/2007-2013) under grant agreement no. 282703.

\end{acknowledgements}

\bibliographystyle{plainnat}

\bibliography{biblio} 

\end{document}